\documentclass[%
 reprint,
aps,
nofootinbib]{revtex4-1}
\usepackage{scrextend}
\usepackage{graphicx}% Include figure files
\usepackage{dcolumn}% Align table columns on decimal point
\usepackage{bm}% bold math
\usepackage{float}
\usepackage{hyperref}% add hypertext capabilities
\usepackage[mathlines]{lineno}% Enable numbering of text and display math
\usepackage{footnote}
\usepackage{amsmath}
\usepackage{xfrac}
\usepackage{braket}
\usepackage{cancel}
\usepackage{xcolor}
\usepackage{amssymb}
\usepackage[mathscr]{euscript}
\usepackage{lipsum}
\usepackage{mathtools}
\usepackage{cuted}
\usepackage[export]{adjustbox}

\usepackage{mathtools}

\newcommand{\arXiv}[2]{\href{http://arxiv.org/pdf/hep-th/#1}{{\tt #2/#1}}}
\newcommand{\arXivpheno}[2]{\href{http://arxiv.org/pdf/hep-ph/#1}{{\tt #2/#1}}}
\newcommand{\arXivold}[2]{\href{http://arxiv.org/pdf/#1}{{\tt #2/#1}}}

\usepackage{blindtext} % just for demo
\begin{document}

\preprint{APS/123-QED}

\title{Implications of the Weak Gravity Conjecture on Charge, Kinetic Mixing, the Photon Mass, and More}% Force line breaks with \\
\author{Fayez Abu-Ajamieh}
 %\altaffiliation[Also at ]{Physics Department, UC Davis.}%Lines break automatically or can be forced with \\
 \email{fayezajamieh@iisc.ac.in}
\affiliation{%
Centre for High Energy Physics, Indian Institute of Science, Bangalore 560012, India
}
\author{Nobuchika Okada}
\email{okadan@ua.edu}
\affiliation{
Department of Physics and Astronomy; 
University of Alabama; Tuscaloosa; Alabama 35487; USA}

\author{Sudhir K. Vempati}
\email{vempati@iisc.ac.in}
\affiliation{Centre for High Energy Physics, Indian Institute of Science, Bangalore 560012, India}


\begin{abstract}
We investigate several phenomenological implications of the Weak Gravity Conjecture (WGC). We find that the WGC implies that the SM neutrinos must be electrically neutral, that the electric charge in the SM must be quantized, and that the photon must be massless. In addition, we use the WGC to set lower bounds on the electric charge of milli-charged particles (mCP), the gauge coupling of several $U(1)$ extensions of the SM, their kinetic mixing parameter with the SM $U(1)_{\text{EM}}$, and the axion couplings to photons and fermions. We also set an upper bound on the lifetime of the proton.

%\begin{description}
%\end{description}
\end{abstract}
\pacs{Valid PACS appear here}% PACS, the Physics and Astronomy
\maketitle

%\tableofcontents

\section{Introduction}\label{sec1}
One of the fundamental questions in Nature is  why gravity is the weakest force amongst the four fundamental interactions. The Weak Gravity Conjecture (WGC)  \cite{Arkani-Hamed:2006emk}
 tries to address the question and  has attracted a lot of attention in recent years. 
A major implication of the conjecture is that any $U(1)$ gauge theory with a coupling $g$, must breakdown at a scale $\Lambda$ below the Planck scale, such that $\Lambda \sim g M_{\text{Pl}}$, where $M_{\text{Pl}} = 2.4 \times 10^{18}$ GeV.
More concretely, consider a $U(1)$ gauge group with coupling $g$. The electric WGC implies that there must be a light charged particle with mass
\begin{equation}\label{eq:Electric_WGC}
m_{\text{el}} \lesssim g_{\text{el}}M_{\text{Pl}},
\end{equation}
which also should hold for magnetic monopoles, i.e.
\begin{equation}\label{eq:Mag_monopole}
m_{\text{mag}} \lesssim g_{\text{mag}} M_{\text{Pl}} \sim \frac{1}{g_{\text{el}}} M_{\text{Pl}}.
\end{equation}

As the monopole has a mass that is at least of the order of the magnetic field it generates, which is linearly divergent, then one has
\begin{equation}\label{eq:Monopole_mass}
m_{\text{mag}} \sim \frac{\Lambda}{g_{\text{el}}^{2}},
\end{equation}
plugging eq. (\ref{eq:Monopole_mass}) in eq. (\ref{eq:Mag_monopole}), we thus arrive at the magnetic WGC
\begin{equation}\label{eq:Magnetic_WGC}
\Lambda \lesssim g M_{\text{Pl}}.
\end{equation}

This result implies that any $U(1)$ gauge theory has a natural cutoff where it breaks down. For instance, for $U(1)_{\text{EM}}$, the WGC suggests $\Lambda \sim 10^{17}$ GeV, close to the scale of the heterotic string. We shall see how eq. (\ref{eq:Magnetic_WGC}) arises naturally when applying the WGC in its simplest form.

A major motivation for this conjecture lies with the fact that the existence of a small charge $g$ without a corresponding light mass $m$ as dictated by eq. (\ref{eq:Electric_WGC}), means that black holes carrying such charges cannot evaporate via the Hawking radiation, which in turn will lead to the problematic issue of remnants \cite{Susskind:1995da}.

The result in eq. (\ref{eq:Magnetic_WGC}) is quite powerful, as instead of predicting a cutoff scale for a $U(1)$ theory, it can be used to set a lower bound on $g$ if the cutoff scale $\Lambda$ is known. In this paper, we shall show that the magnetic WGC has several deep phenomenological implications on various $U(1)$ gauge groups, including the electric neutrality of neutrinos and the electric charge quantization of the SM; setting lower bounds on the electric charge of milli-Charged Particles (mCPs), the charges of $U(1)$ extension of the SM and the kinetic mixing parameter between $U(1)$ and $U(1)_{\text{EM}}$; the masslessness of the photon, and setting lower bounds on the axion's couplings to photons and fermions and an upper bound on the proton's lifetime.

This paper is organized as follows: In Section II, we illustrate the application of the (magnetic) WGC and how to extract the cutoff scale $\Lambda$ to be used with it. In Section III we investigate the various phenomenological implications of the WGC, and then we present our conclusions in Section IV. We relegate some technical details to the Appendix.
 
\section{Applying the WGC}\label{sec2}
\begin{figure}[!t] 
\centering
\includegraphics[width=0.4\textwidth]{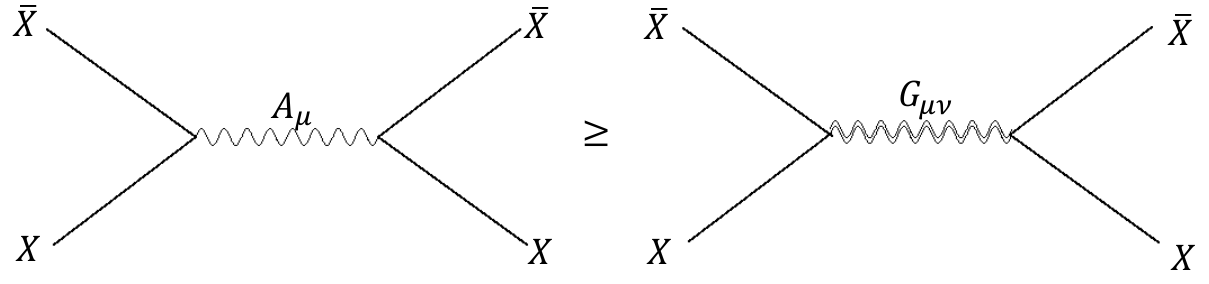}
\caption{Applying the WGC.}
\label{fig1}
\end{figure}

Here we show how to apply the WGC to extract a lower limit on the coupling $g$ of a $U(1)$ gauge group. Consider the s-channel scattering $X\overline{X} \rightarrow X\overline{X}$. Gravitational interaction proceeds through the exchange of a graviton, whereas the $U(1)$ interaction proceeds through the exchange of the corresponding gauge boson. The WGC implies that the gravitational interaction should be weaker than the $U(1)$ interaction, as shown in Figure \ref{fig1}. This implies that
\begin{equation}\label{eq:WGC_general}
|\mathcal{M}_{\text{grav}}(s)| \lesssim |\mathcal{M}_{\text{U(1)}}(s)|.
\end{equation}

Since at high energy $\mathcal{M}_{\text{U(1)}} \sim g^{2}$ and $\mathcal{M}_{\text{grav}} \sim s/M_{\text{Pl}}^{2}$, and if we set set $\sqrt{s} = \Lambda$ as the cutoff scale, then we immediately arrive at eq. (\ref{eq:Magnetic_WGC}).

Notice that if $\Lambda$ is known, or at least if a lower bound on it is known, then one can set a lower bound on $g$. Strictly speaking, $\Lambda$ is interpreted as the scale at which the corresponding QFT breaks down. In our calculation, we will be mainly concerned with the coupling of $U(1)$ to SM fermions and photons. For the former, we can extract a lower limit on $\Lambda$ from the results of LEP, whereas for the latter, we can use the results of Light-By-Light (LBL) scattering in the LHC.

\subsection{The Scale of New Physics (NP) from LEP}\label{sec:2.1}
LEP applied the technique presented in \cite{Eichten:1983hw} on the differential cross-section of $e^{+}e^{-} \rightarrow l^{+}l^{-}, q^{+}q^{-}$ ($l = e, \mu, \tau$ and $q = u,d$); in order to set limits on the scale of NP arising from the operator \cite{ALEPH:2013dgf}
\begin{equation}\label{eq:LEP_Lag}
\mathcal{L}_{\text{eff}} = \frac{g^{2}}{(1+\delta)\Lambda_{\pm}^{2}}\sum_{i,j = L,R} \eta_{ij}(\bar{e}_{i}\gamma_{\mu}e_{i})(\bar{f}_{j}\gamma^{\mu}f_{j}),
\end{equation}
where $\delta = 1(0)$ for $ f = e$ $(f\neq e)$ respectively, $\eta_{ij} \pm 1,0$, $\Lambda_{+}(\Lambda_{-})$ for constructive (destructive) interference with the SM, and by convention $g^{2}/4\pi \rightarrow 1$.

Defining $\epsilon_{\pm} \equiv 1/\Lambda_{\pm}^{2}$ with $\epsilon = 0$ corresponding to the SM, the limits on $\epsilon_{\pm}$ were used to set a $95\%$ C.L. limit on $\Lambda_{\pm}$ of $\sim 10$ TeV. This scale is defined as the lower scale where QFT breaks down if the NP is heavy such that it can be integrated out, i.e., if the NP is of the decoupling type, then we can set $\Lambda \equiv \Lambda_{\text{LEP}} \simeq 10$ TeV. On the other hand, if the NP is light (e.g. axions), then this scale cannot be used as the NP in non-decoupling. Instead, we can drop the corresponding mass and set $\Lambda = \sqrt{s_{\text{LEP}}} = 209$ GeV as the lowest scale where QFT breaks down. 

\subsection{The Scale of NP from Light-By-Light (LBL) Scattering}\label{sec:2.2}
The LEP bound extracted above cannot be used for certain processes when the scattering particles are photons. Instead, the experimental results from LBL scattering should be used. Here, we follow the treatment in \cite{Ellis:2017edi,Ellis:2022uxv}. The effective Lagrangian for LBL scattering can be expressed as:
\begin{equation}\label{eq:LBL_lag}
\mathcal{L}_{\text{eff}} \supset c_{1}F_{\mu\nu}F^{\mu\nu}F_{\rho\sigma}F^{\rho\sigma} + c_{2}F_{\mu\nu}F^{\nu\rho}F_{\rho\sigma}F^{\mu\sigma},
\end{equation}
from which the differential cross-section of LBL scattering at LO is obtained
\begin{equation}\label{eq:diff_Xsection}
\frac{d\sigma}{d\Omega} = \frac{1}{16\pi^{2} \hat{s}}(\hat{s}^{2}+\hat{t}^{2}+\hat{s}\hat{t})^{2}(48 c_{1}^{2} + 11 c_{2}^{2} + 40c_{1}c_{2}),
\end{equation}
where $\hat{s} = m_{\gamma\gamma}^{2}$ and $\hat{t} = -\frac{\hat{s}}{2}(1-\cos{\theta})$. Eq. (\ref{eq:diff_Xsection}) can be immediately integrated to yield the total cross-section
\begin{equation}\label{eq:tot_Xsection}
\sigma_{\gamma\gamma} = \frac{7\hat{s}^{3}}{40\pi}(48 c_{1}^{2} + 11 c_{2}^{2} + 40c_{1}c_{2}).
\end{equation}

Notice that $c_{1,2}$ are model-dependent parameters that have the dimension of $1/\text{(mass)}^{4}$. For example, for the Born-Infeld (BI) model, $c_{1} = -1/32\beta^{2}$ and $c_{2} = 1/8\beta^{2}$, where $\beta$ is the BI scale parameter. To set a conservative limit, we follow the scaling of LEP above and set
\begin{equation}\label{eq:scaling}
c_{1,2} \equiv \frac{g_{i}^{4}}{\Lambda^{4}} \rightarrow \frac{(4\pi)^{2}}{\Lambda^{4}}.
\end{equation}

The latest experimental results of the LBL scattering from ATLAS are given by \cite{ATLAS:2017fur}
\begin{equation}\label{eq:ATLAS}
\sigma_{\gamma\gamma}^{\text{Exp}} = 70 \pm 24 \hspace{1mm} (\text{stat.}) \pm 17 \hspace{1mm} (\text{sys.}) \hspace{1mm} \text{nb};
\end{equation}
setting the theoretical uncertainty at $10$ nb as in \cite{Ellis:2017edi} and summing the uncertainties in quadrature, we get $\delta \sigma^{\text{Exp}} = 31.0644$ nb. Demanding that the cross-section in eq. (\ref{eq:tot_Xsection}) be less than twice this experimental uncertainty and using the scaling in eq. (\ref{eq:scaling}), we can estimate the scale of NP in LBL processes @ $2\sigma$ C.L. as
\begin{equation}\label{eq:LBL_bound}
\Lambda_{\text{LBL}} \gtrsim \Big( \frac{22176\pi^{3}\hat{s}^3}{10\delta \sigma^{\text{Exp}}} \Big)^{1/8} \simeq 7.8 \hspace{1mm} \text{TeV},
\end{equation}
where we have used $\sqrt{\hat{s}} = m_{\gamma\gamma} = 5.02$ TeV \cite{ATLAS:2017fur}.

\section{Phenomenological Implications of the WGC}\label{sec3}
In this section, we discuss several phenomenological implications of the WGC based mainly on eqs. (\ref{eq:Magnetic_WGC}) and (\ref{eq:WGC_general}).
\subsection{Bounds on the Electric Charge}\label{3A}
The WGC can be used to set a lower bound on the electric charge. It is well-known that milli-charges can be introduced to the SM in a variety of ways, such as adding an $SU(2) \times SU(2)$ Dirac fermion with hypercharge $2\epsilon$, \cite{Okun:1983vw}, an additional $U(1)$ \cite{Holdom:1985ag}, allowing neutrinos to have millicharge \cite{Foot:1990uf, Foot:1992ui}, or through introducing non-locality \cite{Capolupo:2022awe, Abu-Ajamieh:2023roj}. For an electric milli-charge $\epsilon e$, the WGC in eq. (\ref{eq:Magnetic_WGC}) sets the lower bound $|\epsilon| \gtrsim \Lambda/(e M_{\text{Pl}})$, which using $\Lambda_{\text{LEP}} = 10$ TeV translates to $|\epsilon| \gtrsim 1.4 \times 10^{-14}$. On the other hand, globular cluster star cooling sets an upper bound of $|\epsilon| \lesssim 2 \times 10^{-14}$ \cite{Davidson:2000hf}. Thus we have
\begin{equation}
1.4 \times 10^{-14} \lesssim |\epsilon| \lesssim 2 \times 10^{-14}.
\end{equation}

For the particular case of SM neutrinos, it can be shown that if they are Dirac fermions, then their charges must be equal \cite{Foot:1990uf, Foot:1992ui, Das:2020egb}
\begin{equation}\label{eq:nu_charges}
Q_{\nu_{e}} = Q_{\nu_{\mu}} = Q_{\nu_{\tau}}.
\end{equation}

Matter neutrality sets an upper bound on their electric charge of  \cite{Marinelli:1983nd,Raffelt:1999gv,Giunti:2014ixa,Bressi:2011yfa, Das:2020egb}
\begin{equation}\label{eq:nu_bound}
|Q_{\nu_{e}}| \lesssim 3.2 \times 10^{-21}e, 
\end{equation}
which in conjunction with the bound from the WGC, and in light of eq. (\ref{eq:nu_charges}) implies that all	 neutrinos must be electrically neutral. Since in the SM model, charge cannot de dequantized without inducing milli-charge in neutrinos \cite{Foot:1990uf, Foot:1992ui, Das:2020egb}, this also implies that in the SM, charge must be quantized.\footnote{If neutrinos are Majorana fermions, then they must be neutral \cite{Babu:1989tq, Babu:1989ex}, which also implies that charge is quantized in the SM.} Thus, we arrive at the following important result: \\

\textit{The WGC, in conjunction with experimental limits, implies that the electric charge in the SM is quantized}.\\

Before we conclude this section, we point out that we can use eq. (\ref{eq:Mag_monopole}) to set a lower bound on the magnetic charge of neutrinos (if such a charge exists)
\begin{equation}\label{eq:nu_magnetic_carge1}
Q_{\nu_{\text{mag}}} \gtrsim \frac{m_{\nu}}{M_{\text{Pl}}} \simeq 4.2 \times 10^{-29},
\end{equation}
where we've used $m_{\nu} = 0.1$ eV. On the other hand, an upper bound on the magnetic charge of neutrinos can be obtained from magnetars \cite{Hook:2017vyc}. Both bounds imply
\begin{equation}\label{eq:nu_magnetic_carge2}
4.2 \times 10^{-29} \lesssim Q_{\nu_{\text{mag}}} \lesssim 10^{-16}.
\end{equation}
\subsection{Implication for Gauged $U(1)$ Extensions}\label{3B}
The WGC can also be used to set bounds on additional gauged $U(1)$ interactions. In particular, a gauged $U(1)$ has a corresponding gauge boson $Z'$, and the process $e^{+}e^{-} \rightarrow Z' \rightarrow \overline{f}f$ yields an effective interaction identical to the one in eq. (\ref{eq:LEP_Lag}). Thus, the LEP results can be used in conjunction with the WGC to set lower bounds on their gauge coupling $g'$. 

The only non-anomalous $U(1)$ gauge group extensions are $B-L$, $L_{\mu}-L_{e}$, $L_{\tau}-L_{e}$ and $L_{\tau}-L_{\mu}$, and we can see that the LEP bound can be applied to the first 3, whereas the LEP bound cannot be used to bound the last as it doesn't interact with electrons at tree-level. The interaction Lagrangians can be expressed as
\begin{eqnarray}\label{eq:U(1)_Lag}
\mathcal{L}_{B-L} & \supset&  g' Z'_{\mu}\Big( \frac{1}{3}(\overline{u}\gamma^{\mu}u +\overline{d}\gamma^{\mu}d) - \overline{e}\gamma^{\mu}e - \overline{\mu}\gamma^{\mu}\mu - \overline{\tau}\gamma^{\mu}\tau\Big), \nonumber \\
\mathcal{L}_{\mu-e} & \supset&  - g' Z'_{\mu}\Big(\overline{e}\gamma^{\mu}e - \overline{\mu}\gamma^{\mu}\mu \Big), \nonumber  \\
\mathcal{L}_{\tau-e} & \supset&  - g' Z'_{\mu}\Big(\overline{e}\gamma^{\mu}e - \overline{\tau}\gamma^{\mu}\tau \Big). 
\end{eqnarray}

For $m_{Z'} \gg \sqrt{s_{\text{LEP}}} = 209$ GeV, $Z'$ can be integrated out to yield an operator identical to eq. (\ref{eq:LEP_Lag}) with $m_{Z'} = \Lambda_{\text{LEP}} = 10$ TeV. Thus, the WGC implies
\begin{equation}\label{eq:HeavyZ1}
\frac{s}{M_{\text{Pl}}^{2}} \lesssim \frac{g'^{2}s}{m_{Z'}^{2}},
\end{equation}
which for $m_{Z'} = 10$ TeV yields the bound
\begin{equation}\label{eq:HeavyZ2}
g' \gtrsim 4.2 \times 10^{-15}.
\end{equation}

On the other hand, if $m_{Z'} \ll \sqrt{s_{\text{LEP}}}$, then it can be neglected to yield an operator identical to eq. (\ref{eq:LEP_Lag}) with $\Lambda = \sqrt{s_{\text{LEP}}}$. In this case the WGC implies
\begin{equation}\label{eq:LightZ1}
g' \gtrsim \frac{\sqrt{s_{\text{LEP}}}}{M_{\text{Pl}}} \simeq 8.7 \times 10^{-17}.
\end{equation}

We can use the experimental bounds on the $g' - m_{Z'}$ parameter space to recast this bound to become a bound on $m_{Z'}$. For $B-L$, the limits in \cite{Heeck:2014zfa} can be used to translate the bound in eq. (\ref{eq:LightZ1}) to become
\begin{equation}\label{eq:Zprime_bound}
m_{Z'} \gtrsim 10^{-2} \hspace{1mm} \text{eV},
\end{equation}
which arises from the cooling of Horizontal Branch (HB) stars, whereas the latest bounds on $L_{\mu} - L_{e}$ and $L_{\tau} - L_{e}$ from \cite{Bauer:2018onh} indicate that no lower bound on the mass can be set.

Another important result is that the bounds in eqs. (\ref{eq:HeavyZ2}) and (\ref{eq:LightZ1}) exclude the viability of an extra $U(1)$ as a solution to the hierarchy problem. In \cite{Craig:2019fdy}, it was proposed to use an extra $U(1)$ gauge group in conjunction with the WGC to furnish a solution to the hierarchy problem. The argument rested on the electric form of the WGC, which postulates the existence of a light particle with mass given by eq. (\ref{eq:Electric_WGC}). More specifically, if neutrinos are Dirac fermions, then they obtain their mass from the Higgs VEV
\begin{equation}\label{eq:nu_mass}
m_{\nu} \sim y_{\nu} v, 
\end{equation}
which according to the WGC should be $m_{\nu} \lesssim g M_{\text{Pl}}$ with $g$ being the coupling of some extra $U(1)$ gauge group. This implies that
\begin{equation}\label{eq:WGC_hierarchy}
\frac{v}{M_{\text{Pl}}} \lesssim \frac{g}{y_{\nu}},
\end{equation}
and a large hierarchy between $v$ and $M_{\text{Pl}}$ can be obtained if $g \ll y_{\nu}$. However, if we use $y_{\nu} \sim 10^{-12}$, then generating the hierarchy between the EW scale and the Planck scale would require
\begin{equation}\label{eq:g_for_hierarchy}
 g \sim \frac{(10^{-12})(246)}{2.4\times 10^{18}} \sim 10^{-28},
\end{equation}
which is many orders of magnitudes below the limits in eqs. (\ref{eq:HeavyZ2}) and (\ref{eq:LightZ1}). This excludes the viability of this solution to the hierarchy problem.\footnote{Although the bounds in eqs. (\ref{eq:HeavyZ2}) and (\ref{eq:LightZ1}) do not hold for the gauge group $L_{\tau}-L_{\mu}$ since the LEP bound does not apply here, using any reasonably low QED cutoff, (even $\sim 1$ GeV), would exclude the viability of this gauge group as well as a solution to the hierarchy problem.}

\subsection{Bounds on Kinetic Mixing}
In addition to setting bounds on the gauge coupling of any additional $U(1)$ gauge group, the WGC can also set bounds on the kinetic mixing parameter of an extra $U(1)$ with the SM $U(1)_{\text{EM}}$. In general, the kinetic term of $U(1) \times U(1)_{\text{EM}}$ can be written as
\begin{equation}\label{eq:KineticMixing}
\mathcal{L}_{\text{kin}} = -\frac{1}{4}F_{\mu\nu}F^{\mu\nu} -\frac{1}{4}X_{\mu\nu}X^{\mu\nu} -\frac{1}{2}\chi F_{\mu\nu}X^{\mu\nu}.
\end{equation}
\begin{figure}[!t] 
\centering
\includegraphics[width=0.3\textwidth]{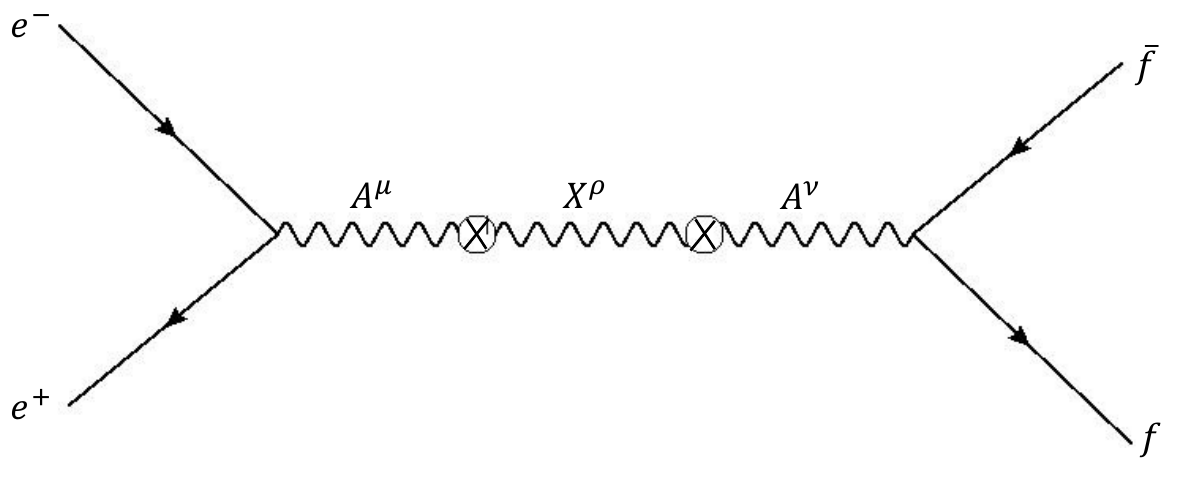}
\caption{$e^{+}e^{-} \rightarrow \overline{f}f$ through kinetic mixing.}
\label{fig2}
\end{figure}
To apply the WGC with the LEP bound, we need two insertions of the kinetic mixing, as shown in Figure (\ref{fig2}). Like the case with $Z'$, the amplitude and bound depends on whether the gauge boson $X$ is light or heavy. 

For $m_{X} \ll \sqrt{s_{\text{LEP}}}$, the amplitude of this process is given by $\mathcal{M} \sim e^{2}\chi^{2}$, which according to the WGC sets the following lower bound on the kinetic mixing parameter
\begin{equation}\label{eq:KinMixingBound_light1}
\chi \gtrsim \frac{\sqrt{s_{\text{LEP}}}}{e M_{\text{Pl}}} \simeq 2.9 \times 10^{-16}.
\end{equation}

Other experimental bounds on the kinetic mixing parameters can be found in \cite{Ahlers:2008qc}. For gauge boson masses up to $\sim 10$ keV , the strongest bounds arise from  the cooling of Red Giant (RG) stars, which is equal to $\sim 6 \times 10^{-14}$. Thus, we can set the following bound
\begin{equation}\label{eq:KinMixingBound_light2}
2.9 \times 10^{-16} \lesssim \chi \lesssim 6 \times 10^{-14} \hspace{2mm} \text{for} \hspace{2mm} m_{X} \lesssim 10 \hspace{1mm} \text{keV}.
\end{equation}

On the other hand, for $m_{X} \gg \sqrt{s_{\text{LEP}}}$, the amplitude is give by $ \mathcal{M} \sim e^{2}\chi^{2}s/m_{X}^{2}$, which according to the WGC sets the following lower bound on the kinetic mixing parameter
\begin{equation}\label{eq:KinMixingBound_heavy1}
\chi \gtrsim \frac{\Lambda}{e M_{\text{Pl}}} \simeq 1.4 \times 10^{-14},
\end{equation}
where $\Lambda = m_{X} = 10$ TeV. For $ m_{X} = 10$ TeV, there is no upper bond on $\chi$. For an earlier analysis using magnetic form of the WGC, please see \cite{Benakli:2020vng}.

\subsection{Bound on Axion Couplings}\label{sec:Axion}
The WGC can also be used to set lower bounds on several axion couplings. The interaction Lagrangian of axions with photons and fermions can be written as
\begin{equation}\label{eq:AxionLag}
\mathcal{L}_{\text{int}} = -\frac{1}{4}g_{a\gamma}a F_{\mu\nu}\tilde{F}^{\mu\nu} + \frac{g_{af}}{2m_{f}}(\partial_{\mu}a)(\overline{f}\gamma^{\mu}\gamma^{5}f),
\end{equation} 
where $\tilde{F}^{\mu\nu} = \frac{1}{2}\epsilon^{\mu\nu\rho\sigma}F_{\rho\sigma}$. The WGC can be used to set bounds on $g_{a\gamma}$ by comparing the amplitude of the LBL scattering through the exchange of an axion with that through the exchange of a graviton. The former is given by $\mathcal{M}_{\text{LBL}(a)} \sim g_{a\gamma}^{2} s$, whereas the latter is given by $\mathcal{M}_{\text{LBL}(G)} \sim s/M_{\text{Pl}}^{2}$. This implies that
\begin{equation}
	g_{g\gamma} \gtrsim \frac{1}{M_{\text{Pl}}},
\end{equation}
which is reminiscent of the Axion Weak Gravity Conjecture \cite{Rudelius:2015xta, Montero:2015ofa, Brown:2015iha, Heidenreich:2015wga, delaFuente:2014aca, Hebecker:2015rya, Bachlechner:2015qja, Rudelius:2014wla, Junghans:2015hba, Kooner:2015rza, Ibanez:2015fcv, Hebecker:2015zss, Hebecker:2019vyf, Daus:2020vtf}. Things are more interesting when we consider the axion's coupling to fermions. If we consider the process $e^{+} e^{-} \rightarrow a \rightarrow \overline{f}f$, then we can use the WGC in conjunction with the LEP results to set a lower bound on the couplings $g_{af}$. Notice here that since axions are light, then the cutoff scale we use is $\Lambda = \sqrt{s_{\text{LEP}}}$. The amplitude of the process is given by $\mathcal{M} \sim g_{ae}g_{af}$, which implies that
\begin{equation}\label{eq:g_aef_bound}
 g_{ae}g_{af} \gtrsim \frac{s_{\text{LEP}}^{2}}{M_{\text{Pl}}^{2}}.
\end{equation} 

For the particular case of $f=e$, this bound reads $g_{ae} \gtrsim 8.7 \times 10^{-17}$. This, together with the bound from cooling of RG stars \cite{Capozzi:2020cbu} implies that 
\begin{equation}\label{eq:g_ae_bound}
8.7 \times 10^{-17} \lesssim g_{ae} \lesssim 1.3 \times 10^{-13}.
\end{equation}

For $f = \mu, \tau, u, d$, the lower limit on $g_{ae}$ can be used in eq. (\ref{eq:g_aef_bound}) to set a conservative lower bound on $g_{af}$
\begin{equation}\label{eq:g_ae_bound}
g_{af} \gtrsim \frac{(209)^{2}}{(2.4\times 10^{18})^{2}(8.7\times 10^{-17})} \simeq 8.7\times 10^{-17}.
\end{equation}

This bound also applies to the axion's coupling to protons and neutrons, since the LEP bound also applies to $e^{+}e^{-} \rightarrow \overline{u}u/\overline{d}d$. Combined with the upper limits on $g_{ap}$ and $g_{an}$ from the cooling of supernovae \cite{Buschmann:2021juv}, the bounds read
\begin{eqnarray}\label{eq:g_aN_bound}
8.7\times 10^{-17} \lesssim g_{ap} \lesssim 1.3 \times 10^{-9}, \\
8.7\times 10^{-17} \lesssim g_{an} \lesssim 1.5 \times 10^{-9}.
\end{eqnarray}

\subsection{Implications of the WGC on the Mass of the Photon}\label{sec:Photon_mass}
Although in the SM the photon is massless, it is possible to give it a small mass while still remaining within experimental bounds. In general, a massless $U(1)$ gauge boson can acquire mass in a gauge-invariant way either via the Stueckelberg mechanism or the Higgs mechanism,\footnote{It is possible for the photon to acquire mass through the Proca action, however, this action is not gauge-invariant.} and for the latter case, the photon can be Higgsed either by the SM Higgs or by a Beyond the SM (BSM) Higgs. Below, we discuss the implication of the WGC for each case individually.

\subsubsection{Photon Mass via the Stueckelberg Mechanism}
The photon can acquire mass via the gauge-invariant Stueckelberg action
\begin{eqnarray}\label{eq:Stueck_Lag}
\mathcal{L}_{\text{Stueck}} & = &  -\frac{1}{2}F_{\mu\nu}^{\dagger}F^{\mu\nu}   - (\partial^{\mu}A_{\mu}^{\dagger} + m B^{\dagger})(\partial^{\nu}A_{\nu} + m B) \nonumber \\ 
& + & m^{2} \Big(A_{\mu}^{\dagger} -\frac{1}{m}\partial_{\mu}B^{\dagger} \Big) \Big(A^{\mu}-\frac{1}{m}\partial^{\mu}B \Big).
\end{eqnarray}

A simple calculation shows that the photon will acquire the following mass \cite{Kuzmin:2001pg,Ruegg:2003ps}
\begin{eqnarray}\label{eq:Stueck_mass}
m_{A}^{2} & = & \frac{1}{8}(g^{2} + g'^{2})v^{2}\Big[ 1 + \epsilon -\sqrt{1-2\epsilon \Big( \frac{g^{2}-g'^{2}}{g^{2}+g'^{2}}\Big) + \epsilon^{2} }\Big], \nonumber \\
 & \simeq & m^{2} \cos^{2}{\theta_{W}},
\end{eqnarray}
where $\theta_{W}$ is the Weinberg angle, and
\begin{equation}
\epsilon = \frac{4m^{2}}{v^{2}(g^{2}+g'^{2})} \ll 1.
\end{equation}

A consequence of the photon acquiring a Stueckelberg mass is that it will interact with neutrinos at tree-level, i.e., neutrinos will acquire an electric charge
\begin{equation}\label{eq:nu_E_coupling}
Q_{\nu} = -\frac{1}{2}g' \cos{\theta_{W}} + \frac{1}{2}g \sin{\theta_{W}} \simeq \frac{g \sin{\theta_{W}}m_{A}^{2}}{2M_{W}^{2}},
\end{equation}
where $M_{W}$ is the mass of the $W$ boson. Given that the upper limit on the photon mass reads \cite{Wu:2016brq, Bonetti:2016cpo,Bonetti:2017pym}
\begin{equation}\label{eq:Photon_mass_limit}
m_{A} \lesssim 1.8 \times 10^{-14} \hspace{2mm} \text{eV},\footnote{This limit is model-independent. A stronger limit of $\sim 10^{-18}$ eV can be obtained with some assumptions \cite{Goldhaber:2008xy, Tu:2005ge}.}
\end{equation}
it is easy to show that the corresponding induced neutrino charge is excluded by the limit in eq. (\ref{eq:nu_bound}). Put differently, we have shown in Section \ref{3A} that the WGC implies that neutrinos must be electrically neutral. Therefore, this implies that the possibility of the photon acquiring mass via the Stueckelberg is excluded.

\subsubsection{Photon Mass via the SM Higgs}
It is also possible for the photon to acquire mass via the SM Higgs mechanism.\footnote{In fact, the Higgs mechanism was originally introduced as a mechanism to give mass to the photon. In this letter, we do not concern ourselves with how Elecroweak Symmetry Breaking (EWSB) is modified when the photon is Higgsed.} In this case, the photon can obtain mass via the kinetic term of the Higgs
\begin{equation}\label{eq:Higgs_kin1}
\mathcal{L}_{\text{SM}}^{A} \supset (D_{\mu}H)^{\dagger} (D^{\mu}H),
\end{equation} 
where $D_{\mu} = \partial_{\mu} + ie Q A_{\mu}$. Inserting the Higgs doublet in the unitary gauge, we can write Lagrangian explicitly as
\begin{eqnarray}\label{eq:Higgs_kin2}
\mathcal{L}_{\text{SM}}^{A} \supset \frac{1}{2} \Big(e^{2} Q^{2} v^{2} + 2 e^{2}Q^{2} v h + e^{2} Q^{2} v^{2}h^{2}\Big)A_{\mu}A^{\mu},
\end{eqnarray} 
and we see now that the photon acquires a mass $m_{A} = e Q v$. Moreover, tree-level cubic and quartic interactions between the Higgs and the photon are induced. 

In order to use the WGC to set a lower limit on the photon mass via the SM Higgs, we can utilize the LBL scattering via the exchange of a Higgs and compare it to that via the exchange of a graviton. For $\sqrt{s} \gg m_{h}$, the former yields $\mathcal{M}_{\text{LBL}} \sim m_{A}^{4}/v^{2} s$, which when compared to the graviton exchange yields the limit
\begin{equation}\label{eq:SM_Higgs_limit}
m_{A} \gtrsim \Lambda_{\text{LBL}} \sqrt{\frac{v}{M_{\text{{Pl}}}}} \simeq 79 \hspace{2mm} \text{keV},
\end{equation} 
which is clearly excluded. Thus, we conclude that the WGC excludes the possibility of a photon mass via the SM Higgs.

\subsubsection{Photon Mass via a BSM Higgs}
It is also possible for the photon to acquire a mass via a BSM Higgs $\phi$. This case is identical to the SM case, except now the VEV $v_{\phi}$ is a free parameter.\footnote{Obviously a BSM Higgs will mix with the SM Higgs at the level of the potential, however, we can set the mixing to be very small to avoid any experimental constraints, since we are only interested in giving mass to the photon.}

Applying the WGC yields a condition similar to eq. (\ref{eq:SM_Higgs_limit}) with $v_{\text{SM}} \rightarrow v_{\phi}$. As such, it is not possible to set a limit on the photon mass directly. However, one can recast eq. (\ref{eq:SM_Higgs_limit}) to set a lower bound on the coupling of $\phi$ to photons by setting
\begin{equation}\label{eq:phi_AA}
\frac{2m_{A}^{2}}{v_{\phi}} \equiv g_{\phi \gamma} \gtrsim \frac{2\Lambda_{\text{LBL}}}{M_{\text{Pl}}} \simeq 5.1 \times 10^{-11} \hspace{2mm} \text{GeV}.
\end{equation}

On the other hand, one can set an upper bound on $g_{\phi \gamma}$ from the cooling of HB stars (see the Appendix)
\begin{equation}\label{eq:HB_bound}
g_{\phi\gamma} \lesssim 2 \times 10^{-38} \hspace{2mm} \text{GeV},
\end{equation}
which clearly excludes the possibility of a photon mass via a BSM Higgs as well. Thus, the results of this section show that:\\

\textit{The WGC, in conjunction with the experimental limits, implies that the photon is strictly massless}.

\subsection{Implications of the WGC on the Proton Lifetime}
It is well-known that Grand Unified Theories (GUTs) \cite{Pati:1973uk, Georgi:1972cj, Fritzsch:1974nn} lead to proton decay \cite{Weinberg:1979sa, Wilczek:1979hc, Weinberg:1980bf, Weinberg:1981wj, Sakai:1981pk, Dimopoulos:1981dw, Ellis:1981tv} $p^{+} \rightarrow e^{+} \pi^{0}$ due to the exchange of a leptoquark X. The lifetime of the proton decay is given by
\begin{equation}\label{eq:proton_decay}
\tau_{p^{+} \rightarrow e^{+} \pi^{0}} \sim \frac{M_{X}^{4}}{m_{p}^{5}},
\end{equation}
where $M_{X}$ is the mass of the leptoquark and $m_{p}$ is the mass of the proton. Since the X is the gauge boson, the WGC can be applied here as well. The  WGC (eq. (\ref{eq:Electric_WGC})) implies that the upper bound of the mass of X is $M_{\text{Pl}}$. This translates into an upper bound of the lifetime of the proton
\begin{equation}\label{eq:proton_decay}
\tau_{p^{+} \rightarrow e^{+} \pi^{0}} \lesssim \frac{M_{\text{Pl}}^{4}}{m_{p}^{5}} \sim 10^{42} \hspace{2mm}\text{yr.}
\end{equation}

Unfortunately, this bound is much weaker than the current bound of $\sim 10^{36} \hspace{2mm}\text{yr}$ for non-SUSY SU(5) and  $\sim 10^{39} \hspace{2mm}\text{yr}$ for SUSY SU(5) that is calculated in \cite{Dorsner:2005ii, Dorsner:2006hw}.

\section{Conclusions}\label{sec:4}
In this paper, we have shown that the WGC, in conjunction with the relevant experimental limits, can have many important phenomenological implications, including the electric neutrality of neutrinos, charge quantization of the SM, and the masslessness of photons. In addition, we have used the WGC to set lower limits on the electric charge of mCPs, the coupling of several BSM $U(1)$ gauge groups and their kinetic mixing parameter with the SM $U(1)$ gauge group, and the coupling of the axion to photons and fermions. These limits can be used in synergy with the limits from experiment to define the viable range of these couplings that is bound both from above and below. Other set of complementarity bounds can come from theoretical constraints like positivity conditions in the presence of gravity\cite{Tran}. 

The profound implications of the WGC provide motivation for further efforts study it in  more depth, including trying to prove or disprove it.

\section*{Acknowledgments}
The work of FA is supported by the C.V. Raman fellowship from CHEP at IISc. The
work on NO is supported in part by the United States Department of Energy.
\appendix

\section{Setting an Upper Bound on $g_{\phi\gamma}$ from the Cooling of HB Stars}\label{appendix1}
Here we show how we obtain the bound in eq. (\ref{eq:HB_bound}) from the cooling HB stars. But before doing so, let's try to set a limit on $m_{\phi}$. First, notice that we can use the upper limit on the photon mass in (eq. \ref{eq:Photon_mass_limit}) with eq. (\ref{eq:SM_Higgs_limit}) to set an upper bound on the BSM VEV
\begin{equation}\label{eq:VEV_upper_bound}
v_{\phi} \lesssim 1.3 \times 10^{-26} \hspace{2mm} \text{eV},
\end{equation} 
where we have used $\Lambda_{\text{LBL}}$. If we assume that the quartic coupling of $\phi$ is $\lambda_{\phi}$, such that $m_{\phi} = \sqrt{2\lambda_{\phi}}v_{\phi}$, then we can also set an upper bound on $m_{\phi}$ by assuming that $\lambda_{\phi}$ saturates the perturbativity limit of $4\pi$. Thus we obtain
\begin{equation}\label{eq:mass_upper_bound}
m_{\phi} \lesssim 6.4 \times 10^{-26} \hspace{2mm} \text{eV}.
\end{equation}

\begin{figure}[!h] 
\centering
\includegraphics[width=0.3\textwidth]{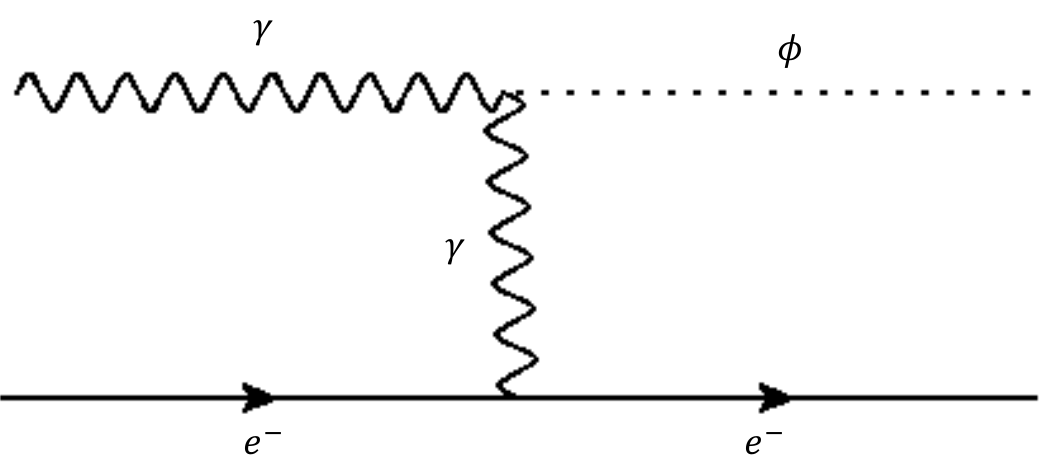}
\caption{Primakoff production.}
\label{fig3}
\end{figure}

Thus, we see that any BSM Higgs that gives mass of the photon must be extremely light, justifying the assumption of small mixing with the SM Higgs. The dominant process in HB cooling is the Primakoff (or inverse Primakoff) production shown in Figure \ref{fig3}.

Assuming that the electron is non-relativistic and neglecting its recoil energy, both of which are reasonable assumptions given the conditions of the HB medium and the smallness of $m_{\phi}$, the cross-section of the Primakoff production is given by
\begin{equation}\label{eq:Primakoff}
\sigma(\omega) \simeq \frac{\alpha g_{\phi\gamma}^{2}}{16 m_{e}^{2} \omega^{2}}\Bigg[ \log{\Big( \frac{4\omega^{2}}{m_{A}^{2}+m_{\phi}^{2}}\Big)} + \frac{2m_{e}^{2}}{m_{A}^{2}+m_{\phi}^{2}}\Bigg],
\end{equation}
where $\omega$ is the photon energy. The energy loss rate per unit volume can be expressed as \cite{Raffelt:1996wa}
\begin{equation}\label{eq:E_loss_rate}
Q = 2n_{e} \int \frac{d^{3}\vec{k}}{(2\pi)^{3}} \frac{\omega\sigma(\omega)}{e^{\omega/T}-1},
\end{equation}
where $n_{e}$ is the electron number density and $T$ is the temperature of the star. To set a conservative limit on $g_{\phi\gamma}$, we assume $m_{A}$ to be the largest possible value allowed by experiment (saturating the upper limit in eq. (\ref{eq:Photon_mass_limit})). This implies that $m_{\phi} \ll m_{A}$, so we drop  $m_{\phi}$ in our calculation. Thus, plugging eq. (\ref{eq:Primakoff}) in eq. (\ref{eq:E_loss_rate}), the upper bound on  $g_{\phi\gamma}$ can be expressed as
\begin{equation}
g_{\phi\gamma} \lesssim \sqrt{\frac{8\pi^{2}m_{e}^{2}m_{u} \dot{\varepsilon}_{\text{max}}}{\alpha y_{e}I(\omega)}},
\end{equation}
where $m_{u}$ is the atomic mass unit, $y_{e} = n_{e}m_{u}/\rho$ is the electron fraction in the star, $\dot{\varepsilon}_{\text{max}} = Q/\rho$ is the maximum energy loss rate per unit mass, $\rho$ is the density of the star, and the integral $I(\omega)$ is given by
\begin{equation}\label{eq:Integral}
I(\omega) = \int_{0}^{\infty} d\omega \Big( \frac{\omega}{e^{\omega/T}-1}\Big)\Bigg[ \log{\Big( \frac{2\omega}{m_{A}}\Big)} + \frac{m_{e}^{2}}{m_{A}^{2}} \Bigg].
\end{equation}

For HB stars, $\dot{\varepsilon}_{\text{max}} = 10 \hspace{1mm} \text{erg} \hspace{1mm} \text{s}^{-1} \hspace{1mm} \text{g}^{-1}$. Using $T = 10^{8} \hspace{1mm} \text{K}$, $\rho = 10^{4} \hspace{1mm} \text{g} \hspace{1mm} \text{cm}^{-3}$ and $y_{e} = 0.5$; and evaluating the integral in eq. (\ref{eq:Integral}) numerically, we arrive at the upper bound:
\begin{equation}\label{eq:HB_bound2}
	g_{\phi\gamma} \lesssim 2 \times 10^{-38} \hspace{2mm} \text{GeV}.
\end{equation}

This bound, however, is only valid in the free streaming regime, i.e., when $\phi$ in the Primakoff process, or the photon in the inverse Primakoff process, stream freely out of the star. To verify the validity of this assumption, we need to check both the decay length and the mean free path.

To find the decay length, let's assume that $\phi$ only couples to the photon at tree-level. Given the upper limit on the mass of $\phi$, we can see that it can only decay to a pair of photons if $m_{\phi} \geq 2m_{A}$. Otherwise $\phi$ would be stable. Assuming that $\phi$ can decay to a pair of photons, the decay width is given by
\begin{equation}\label{eq:Decay_width}
\Gamma_{\phi} = \frac{1}{\tau_{\phi}} = \frac{2\lambda_{\phi} m_{A}^{4}}{\pi m_{\phi}^{3}},
\end{equation}
where we have assumed $\phi$ has a quartic coupling $\lambda_{\phi}$. The decay length is given by
\begin{equation}\label{eq:Decay_length}
L_{\phi} = \tau_{\phi}\beta \gamma c,
\end{equation}
where $\gamma  = 1/\sqrt{1-\beta^{2}} = E_{\phi}/m_{\phi}$. Assuming that $\phi$ is relativistic, its average energy is given by
\begin{equation}\label{eq:Average_E_rel}
\langle E_{\phi} \rangle = \frac{\pi^{4}}{30\xi(3)}T \simeq	2.701 T.
\end{equation}

To get a conservative limit, we set $\lambda = 4\pi$ and use the upper limit on $m_{\phi}$. This yields a lower limit on the decay length of $ \sim O(10^{43})$ Km, which is many orders of magnitude larger than the typical radius of an HB star of $\sim 10^{9}$ Km. Thus, $\phi$ does not decay inside the star.

Next, we evaluate the mean free path. The mfp can be expressed as
\begin{equation}\label{eq:mfp}
\text{mfp} = \frac{1}{n_{e} \sigma (\omega)}.
\end{equation}

Using $ \langle \omega \rangle \simeq 2.701 T$, we find that $\text{mph} \simeq 4 \times 10^{18}$ Km, which is also many orders of magnitude larger than radius of an HB star, thereby justifying our free streaming assumption.


\begin{thebibliography}{10}

%\cite{Arkani-Hamed:2006emk}
\bibitem{Arkani-Hamed:2006emk}
N.~Arkani-Hamed, L.~Motl, A.~Nicolis and C.~Vafa,
``The String landscape, black holes and gravity as the weakest force,''
JHEP \textbf{06}, 060 (2007)
\arXiv{0601001}{hep-th}.
%1148 citations counted in INSPIRE as of 17 Sep 2023

%\cite{Susskind:1995da}
\bibitem{Susskind:1995da}
L.~Susskind,
``Trouble for remnants,''
\arXiv{9501106}{hep-th}.
%150 citations counted in INSPIRE as of 24 Sep 2023

%\cite{Eichten:1983hw}
\bibitem{Eichten:1983hw}
E.~Eichten, K.~D.~Lane and M.~E.~Peskin,
``New Tests for Quark and Lepton Substructure,''
Phys. Rev. Lett. \textbf{50}, 811-814 (1983)
%1131 citations counted in INSPIRE as of 17 Sep 2023

%\cite{ALEPH:2013dgf}
\bibitem{ALEPH:2013dgf}
S.~Schael \textit{et al.} [ALEPH, DELPHI, L3, OPAL and LEP Electroweak],
``Electroweak Measurements in Electron-Positron Collisions at W-Boson-Pair Energies at LEP,''
Phys. Rept. \textbf{532}, 119-244 (2013)
\arXivold{1302.3415}{hep-ex}.
%766 citations counted in INSPIRE as of 17 Sep 2023

%\cite{Ellis:2017edi}
\bibitem{Ellis:2017edi}
J.~Ellis, N.~E.~Mavromatos and T.~You,
``Light-by-Light Scattering Constraint on Born-Infeld Theory,''
Phys. Rev. Lett. \textbf{118}, no.26, 261802 (2017)
\arXivold{1703.08450}{hep-ph}.
%117 citations counted in INSPIRE as of 01 Dec 2023

%\cite{Ellis:2022uxv}
\bibitem{Ellis:2022uxv}
J.~Ellis, N.~E.~Mavromatos, P.~Roloff and T.~You,
``Light-by-light scattering at future $e^+e^-$ colliders,''
Eur. Phys. J. C \textbf{82}, no.7, 634 (2022)
\arXivold{2203.17111}{hep-ph}.
%13 citations counted in INSPIRE as of 17 Sep 2023


%\cite{ATLAS:2017fur}
\bibitem{ATLAS:2017fur}
M.~Aaboud \textit{et al.} [ATLAS],
``Evidence for light-by-light scattering in heavy-ion collisions with the ATLAS detector at the LHC,''
Nature Phys. \textbf{13}, no.9, 852-858 (2017)
\arXivold{1702.01625}{hep-ex}.
%313 citations counted in INSPIRE as of 17 Sep 2023

%\cite{Okun:1983vw}
\bibitem{Okun:1983vw}
L.~B.~Okun, M.~B.~Voloshin and V.~I.~Zakharov,
``ELECTRICAL NEUTRALITY OF ATOMS AND GRAND UNIFICATION MODELS,''
Phys. Lett. B \textbf{138}, 115-120 (1984)
%50 citations counted in INSPIRE as of 24 Sep 2023


%\cite{Holdom:1985ag}
\bibitem{Holdom:1985ag}
B.~Holdom,
``Two U(1)'s and Epsilon Charge Shifts,''
Phys. Lett. B \textbf{166}, 196-198 (1986)
%2356 citations counted in INSPIRE as of 24 Sep 2023


%\cite{Foot:1990uf}
\bibitem{Foot:1990uf}
R.~Foot, G.~C.~Joshi, H.~Lew and R.~R.~Volkas,
``Charge quantization in the standard model and some of its extensions,''
Mod. Phys. Lett. A \textbf{5}, 2721-2732 (1990)
%91 citations counted in INSPIRE as of 24 Sep 2023

%\cite{Foot:1992ui}
\bibitem{Foot:1992ui}
R.~Foot, H.~Lew and R.~R.~Volkas,
``Electric charge quantization,''
J. Phys. G \textbf{19}, 361-372 (1993)
[erratum: J. Phys. G \textbf{19}, 1067 (1993)]
\arXivpheno{9209259}{hep-ph}.
%92 citations counted in INSPIRE as of 24 Sep 2023

%\cite{Capolupo:2022awe}
\bibitem{Capolupo:2022awe}
A.~Capolupo, G.~Lambiase and A.~Quaranta,
``Muon $g-2$ anomaly and non-locality,''
Phys. Lett. B \textbf{829}, 137128 (2022)
\arXivold{2206.06037}{hep-ph}.
%7 citations counted in INSPIRE as of 24 Sep 2023

%\cite{Abu-Ajamieh:2023roj}
\bibitem{Abu-Ajamieh:2023roj}
F.~Abu-Ajamieh, P.~Chattopadhyay, A.~Ghoshal and N.~Okada,
``Anomalies in String-inspired Non-local Extensions of QED,''
\arXivold{2307.01589}{hep-th}.
%1 citations counted in INSPIRE as of 24 Sep 2023

%\cite{Davidson:2000hf}
\bibitem{Davidson:2000hf}
S.~Davidson, S.~Hannestad and G.~Raffelt,
``Updated bounds on millicharged particles,''
JHEP \textbf{05}, 003 (2000)
\arXivpheno{0001179}{hep-ph}.
%460 citations counted in INSPIRE as of 24 Sep 2023


%\cite{Marinelli:1983nd}
\bibitem{Marinelli:1983nd}
M.~Marinelli and G.~Morpurgo,
``The Electric Neutrality of Matter: A Summary,''
Phys. Lett. B \textbf{137}, 439-442 (1984)
%91 citations counted in INSPIRE as of 24 Sep 2023

%\cite{Raffelt:1999gv}
\bibitem{Raffelt:1999gv}
G.~G.~Raffelt,
``Limits on neutrino electromagnetic properties: An update,''
Phys. Rept. \textbf{320}, 319-327 (1999)
%175 citations counted in INSPIRE as of 24 Sep 2023

%\cite{Giunti:2014ixa}
\bibitem{Giunti:2014ixa}
C.~Giunti and A.~Studenikin,
``Neutrino electromagnetic interactions: a window to new physics,''
Rev. Mod. Phys. \textbf{87}, 531 (2015)
\arXivold{1403.6344}{hep-ph}.
%365 citations counted in INSPIRE as of 24 Sep 2023

%\cite{Bressi:2011yfa}
\bibitem{Bressi:2011yfa}
G.~Bressi, G.~Carugno, F.~Della Valle, G.~Galeazzi, G.~Ruoso and G.~Sartori,
``Testing the neutrality of matter by acoustic means in a spherical resonator,''
Phys. Rev. A \textbf{83}, no.5, 052101-1-052101-14 (2011)
\arXivold{1102.2766}{physics.atom-ph}.
%26 citations counted in INSPIRE as of 24 Sep 2023

%\cite{Das:2020egb}
\bibitem{Das:2020egb}
A.~Das, D.~Ghosh, C.~Giunti and A.~Thalapillil,
``Neutrino charge constraints from scattering to the weak gravity conjecture to neutron stars,''
Phys. Rev. D \textbf{102}, no.11, 115009 (2020)
\arXivold{2005.12304}{hep-ph}.
%7 citations counted in INSPIRE as of 24 Sep 2023

%\cite{Babu:1989tq}
\bibitem{Babu:1989tq}
K.~S.~Babu and R.~N.~Mohapatra,
``Is There a Connection Between Quantization of Electric Charge and a Majorana Neutrino?,''
Phys. Rev. Lett. \textbf{63}, 938 (1989)
%118 citations counted in INSPIRE as of 24 Sep 2023

%\cite{Babu:1989ex}
\bibitem{Babu:1989ex}
K.~S.~Babu and R.~N.~Mohapatra,
``Quantization of Electric Charge From Anomaly Constraints and a Majorana Neutrino,''
Phys. Rev. D \textbf{41}, 271 (1990)
%112 citations counted in INSPIRE as of 24 Sep 2023

%\cite{Hook:2017vyc}
\bibitem{Hook:2017vyc}
A.~Hook and J.~Huang,
``Bounding millimagnetically charged particles with magnetars,''
Phys. Rev. D \textbf{96}, no.5, 055010 (2017)
\arXivold{1705.01107}{hep-ph}.
%27 citations counted in INSPIRE as of 24 Sep 2023

%\cite{Heeck:2014zfa}
\bibitem{Heeck:2014zfa}
J.~Heeck,
``Unbroken B \textendash{} L symmetry,''
Phys. Lett. B \textbf{739}, 256-262 (2014)
\arXivold{1408.6845}{hep-ph}.
%160 citations counted in INSPIRE as of 24 Sep 2023

%\cite{Bauer:2018onh}
\bibitem{Bauer:2018onh}
M.~Bauer, P.~Foldenauer and J.~Jaeckel,
``Hunting All the Hidden Photons,''
JHEP \textbf{07}, 094 (2018)
\arXivold{1803.05466}{hep-ph}.
%296 citations counted in INSPIRE as of 24 Sep 2023

%\cite{Craig:2019fdy}
\bibitem{Craig:2019fdy}
N.~Craig, I.~Garcia Garcia and S.~Koren,
``The Weak Scale from Weak Gravity,''
JHEP \textbf{09}, 081 (2019)
\arXivold{1904.08426}{hep-ph}.
%24 citations counted in INSPIRE as of 24 Sep 2023

%\cite{Ahlers:2008qc}
\bibitem{Ahlers:2008qc}
M.~Ahlers, J.~Jaeckel, J.~Redondo and A.~Ringwald,
``Probing Hidden Sector Photons through the Higgs Window,''
Phys. Rev. D \textbf{78}, 075005 (2008)
\arXivold{0807.4143}{hep-ph}.
%102 citations counted in INSPIRE as of 24 Sep 2023



%\cite{Benakli:2020vng}
\bibitem{Benakli:2020vng}
K.~Benakli, C.~Branchina and G.~Lafforgue-Marmet,
%``U(1) mixing and the Weak Gravity Conjecture,''
Eur. Phys. J. C \textbf{80} (2020) no.12, 1118
doi:10.1140/epjc/s10052-020-08691-4
[arXiv:2007.02655 [hep-ph]].
%22 citations counted in INSPIRE as of 29 Jan 2024

%\cite{Rudelius:2015xta}
\bibitem{Rudelius:2015xta}
T.~Rudelius,
``Constraints on Axion Inflation from the Weak Gravity Conjecture,''
JCAP \textbf{09}, 020 (2015)
\arXivold{1503.00795}{hep-th}.
%200 citations counted in INSPIRE as of 17 Dec 2023

%\cite{Montero:2015ofa}
\bibitem{Montero:2015ofa}
M.~Montero, A.~M.~Uranga and I.~Valenzuela,
``Transplanckian axions!?,''
JHEP \textbf{08}, 032 (2015)
\arXivold{1503.03886}{hep-th}.
%190 citations counted in INSPIRE as of 17 Dec 2023

%\cite{Brown:2015iha}
\bibitem{Brown:2015iha}
J.~Brown, W.~Cottrell, G.~Shiu and P.~Soler,
``Fencing in the Swampland: Quantum Gravity Constraints on Large Field Inflation,''
JHEP \textbf{10}, 023 (2015)
\arXivold{1503.04783}{hep-th}.
%229 citations counted in INSPIRE as of 17 Dec 2023


%\cite{Heidenreich:2015wga}
\bibitem{Heidenreich:2015wga}
B.~Heidenreich, M.~Reece and T.~Rudelius,
``Weak Gravity Strongly Constrains Large-Field Axion Inflation,''
JHEP \textbf{12}, 108 (2015)
\arXivold{1506.03447}{hep-th}.
%169 citations counted in INSPIRE as of 17 Dec 2023

%\cite{delaFuente:2014aca}
\bibitem{delaFuente:2014aca}
A.~de la Fuente, P.~Saraswat and R.~Sundrum,
``Natural Inflation and Quantum Gravity,''
Phys. Rev. Lett. \textbf{114}, no.15, 151303 (2015)
\arXivold{1412.3457}{hep-th}.
%150 citations counted in INSPIRE as of 17 Dec 2023

%\cite{Hebecker:2015rya}
\bibitem{Hebecker:2015rya}
A.~Hebecker, P.~Mangat, F.~Rompineve and L.~T.~Witkowski,
``Winding out of the Swamp: Evading the Weak Gravity Conjecture with F-term Winding Inflation?,''
Phys. Lett. B \textbf{748}, 455-462 (2015)
\arXivold{1503.07912}{hep-th}.
%136 citations counted in INSPIRE as of 17 Dec 2023

%\cite{Bachlechner:2015qja}
\bibitem{Bachlechner:2015qja}
T.~C.~Bachlechner, C.~Long and L.~McAllister,
``Planckian Axions and the Weak Gravity Conjecture,''
JHEP \textbf{01}, 091 (2016)
\arXivold{1503.07853}{hep-th}.
%142 citations counted in INSPIRE as of 17 Dec 2023

%\cite{Rudelius:2014wla}
\bibitem{Rudelius:2014wla}
T.~Rudelius,
``On the Possibility of Large Axion Moduli Spaces,''
JCAP \textbf{04}, 049 (2015)
\arXivold{1409.5793}{hep-th}.
%121 citations counted in INSPIRE as of 17 Dec 2023

%\cite{Junghans:2015hba}
\bibitem{Junghans:2015hba}
D.~Junghans,
``Large-Field Inflation with Multiple Axions and the Weak Gravity Conjecture,''
JHEP \textbf{02}, 128 (2016)
\arXivold{1504.03566}{hep-th}.
%87 citations counted in INSPIRE as of 17 Dec 2023

%\cite{Kooner:2015rza}
\bibitem{Kooner:2015rza}
K.~Kooner, S.~Parameswaran and I.~Zavala,
``Warping the Weak Gravity Conjecture,''
Phys. Lett. B \textbf{759}, 402-409 (2016)
\arXivold{1509.07049}{hep-th}.
%77 citations counted in INSPIRE as of 17 Dec 2023

%\cite{Ibanez:2015fcv}
\bibitem{Ibanez:2015fcv}
L.~E.~Ibanez, M.~Montero, A.~Uranga and I.~Valenzuela,
``Relaxion Monodromy and the Weak Gravity Conjecture,''
JHEP \textbf{04}, 020 (2016)
\arXivold{1512.00025}{hep-th}.
%145 citations counted in INSPIRE as of 17 Dec 2023

%\cite{Hebecker:2015zss}
\bibitem{Hebecker:2015zss}
A.~Hebecker, F.~Rompineve and A.~Westphal,
``Axion Monodromy and the Weak Gravity Conjecture,''
JHEP \textbf{04}, 157 (2016)
\arXivold{1512.03768}{hep-th}.
%119 citations counted in INSPIRE as of 17 Dec 2023

%\cite{Hebecker:2019vyf}
\bibitem{Hebecker:2019vyf}
A.~Hebecker and P.~Henkenjohann,
``Gauge and gravitational instantons: From 3-forms and fermions to Weak Gravity and flat axion potentials,''
JHEP \textbf{09}, 038 (2019)
\arXivold{1906.07728}{hep-th}.
%16 citations counted in INSPIRE as of 17 Dec 2023

%\cite{Daus:2020vtf}
\bibitem{Daus:2020vtf}
T.~Daus, A.~Hebecker, S.~Leonhardt and J.~March-Russell,
``Towards a Swampland Global Symmetry Conjecture using weak gravity,''
Nucl. Phys. B \textbf{960}, 115167 (2020)
\arXivold{2002.02456}{hep-th}.
%35 citations counted in INSPIRE as of 17 Dec 2023

%\cite{Capozzi:2020cbu}
\bibitem{Capozzi:2020cbu}
F.~Capozzi and G.~Raffelt,
``Axion and neutrino bounds improved with new calibrations of the tip of the red-giant branch using geometric distance determinations,''
Phys. Rev. D \textbf{102}, no.8, 083007 (2020)
\arXivold{2007.03694}{astro-ph.SR}.
%120 citations counted in INSPIRE as of 24 Sep 2023

%\cite{Buschmann:2021juv}
\bibitem{Buschmann:2021juv}
M.~Buschmann, C.~Dessert, J.~W.~Foster, A.~J.~Long and B.~R.~Safdi,
``Upper Limit on the QCD Axion Mass from Isolated Neutron Star Cooling,''
Phys. Rev. Lett. \textbf{128}, no.9, 091102 (2022)
\arXivold{2111.09892}{hep-ph}.
%44 citations counted in INSPIRE as of 24 Sep 2023

%\cite{Kuzmin:2001pg}
\bibitem{Kuzmin:2001pg}
S.~V.~Kuzmin and D.~G.~C.~McKeon,
``Stueckelberg mass in the Glashow-Weinberg-Salam model,''
Mod. Phys. Lett. A \textbf{16}, 747-753 (2001)
%16 citations counted in INSPIRE as of 29 Sep 2023

%\cite{Ruegg:2003ps}
\bibitem{Ruegg:2003ps}
H.~Ruegg and M.~Ruiz-Altaba,
``The Stueckelberg field,''
Int. J. Mod. Phys. A \textbf{19}, 3265-3348 (2004)
\arXiv{0304245}{hep-th}.
%386 citations counted in INSPIRE as of 29 Sep 2023


%\cite{Wu:2016brq}
\bibitem{Wu:2016brq}
X.~F.~Wu, S.~B.~Zhang, H.~Gao, J.~J.~Wei, Y.~C.~Zou, W.~H.~Lei, B.~Zhang, Z.~G.~Dai and P.~M\'esz\'aros,
``Constraints on the Photon Mass with Fast Radio Bursts,''
Astrophys. J. Lett. \textbf{822}, no.1, L15 (2016)
\arXivold{1602.07835}{astro-ph.HE}.
%70 citations counted in INSPIRE as of 29 Sep 2023

%\cite{Bonetti:2016cpo}
\bibitem{Bonetti:2016cpo}
L.~Bonetti, J.~Ellis, N.~E.~Mavromatos, A.~S.~Sakharov, E.~K.~G.~Sarkisyan-Grinbaum and A.~D.~A.~M.~Spallicci,
``Photon Mass Limits from Fast Radio Bursts,''
Phys. Lett. B \textbf{757}, 548-552 (2016)
\arXivold{1602.09135}{astro-ph.HE}.
%65 citations counted in INSPIRE as of 29 Sep 2023

%\cite{Bonetti:2017pym}
\bibitem{Bonetti:2017pym}
L.~Bonetti, J.~Ellis, N.~E.~Mavromatos, A.~S.~Sakharov, E.~K.~Sarkisyan-Grinbaum and A.~D.~A.~M.~Spallicci,
``FRB 121102 Casts New Light on the Photon Mass,''
Phys. Lett. B \textbf{768}, 326-329 (2017)
\arXivold{1701.03097}{astro-ph.HE}.
%44 citations counted in INSPIRE as of 29 Sep 2023


%\cite{Goldhaber:2008xy}
\bibitem{Goldhaber:2008xy}
A.~S.~Goldhaber and M.~M.~Nieto,
``Photon and Graviton Mass Limits,''
Rev. Mod. Phys. \textbf{82}, 939-979 (2010)
\arXivold{0809.1003}{hep-ph}.
%295 citations counted in INSPIRE as of 29 Sep 2023

%\cite{Tu:2005ge}
\bibitem{Tu:2005ge}
L.~C.~Tu, J.~Luo and G.~T.~Gillies,
``The mass of the photon,''
Rept. Prog. Phys. \textbf{68}, 77-130 (2005)
%120 citations counted in INSPIRE as of 29 Sep 2023


%\cite{Pati:1973uk}
\bibitem{Pati:1973uk}
J.~C.~Pati and A.~Salam,
``Unified Lepton-Hadron Symmetry and a Gauge Theory of the Basic Interactions,''
Phys. Rev. D \textbf{8}, 1240-1251 (1973)
%1418 citations counted in INSPIRE as of 17 Nov 2023

%\cite{Georgi:1972cj}
\bibitem{Georgi:1972cj}
H.~Georgi and S.~L.~Glashow,
``Unified weak and electromagnetic interactions without neutral currents,''
Phys. Rev. Lett. \textbf{28}, 1494 (1972)
%592 citations counted in INSPIRE as of 17 Nov 2023

%\cite{Fritzsch:1974nn}
\bibitem{Fritzsch:1974nn}
H.~Fritzsch and P.~Minkowski,
``Unified Interactions of Leptons and Hadrons,''
Annals Phys. \textbf{93}, 193-266 (1975)
%2144 citations counted in INSPIRE as of 17 Nov 2023


%\cite{Weinberg:1979sa}
\bibitem{Weinberg:1979sa}
S.~Weinberg,
``Baryon and Lepton Nonconserving Processes,''
Phys. Rev. Lett. \textbf{43}, 1566-1570 (1979)
%2315 citations counted in INSPIRE as of 17 Nov 2023

%\cite{Wilczek:1979hc}
\bibitem{Wilczek:1979hc}
F.~Wilczek and A.~Zee,
``Operator Analysis of Nucleon Decay,''
Phys. Rev. Lett. \textbf{43}, 1571-1573 (1979)
doi:10.1103/PhysRevLett.43.1571
%610 citations counted in INSPIRE as of 17 Nov 2023

%\cite{Weinberg:1980bf}
\bibitem{Weinberg:1980bf}
S.~Weinberg,
``Varieties of Baryon and Lepton Nonconservation,''
Phys. Rev. D \textbf{22}, 1694 (1980)
%380 citations counted in INSPIRE as of 17 Nov 2023

%\cite{Weinberg:1981wj}
\bibitem{Weinberg:1981wj}
S.~Weinberg,
``Supersymmetry at Ordinary Energies. 1. Masses and Conservation Laws,''
Phys. Rev. D \textbf{26}, 287 (1982)
%1184 citations counted in INSPIRE as of 17 Nov 2023

%\cite{Sakai:1981pk}
\bibitem{Sakai:1981pk}
N.~Sakai and T.~Yanagida,
``Proton Decay in a Class of Supersymmetric Grand Unified Models,''
Nucl. Phys. B \textbf{197}, 533 (1982)
%960 citations counted in INSPIRE as of 17 Nov 2023

%\cite{Dimopoulos:1981dw}
\bibitem{Dimopoulos:1981dw}
S.~Dimopoulos, S.~Raby and F.~Wilczek,
``Proton Decay in Supersymmetric Models,''
Phys. Lett. B \textbf{112}, 133 (1982)
%463 citations counted in INSPIRE as of 17 Nov 2023

%\cite{Ellis:1981tv}
\bibitem{Ellis:1981tv}
J.~R.~Ellis, D.~V.~Nanopoulos and S.~Rudaz,
``GUTs 3: SUSY GUTs 2,''
Nucl. Phys. B \textbf{202}, 43-62 (1982)
%399 citations counted in INSPIRE as of 17 Nov 2023


%\cite{Dorsner:2005ii}
\bibitem{Dorsner:2005ii}
I.~Dorsner, P.~Fileviez Perez and R.~Gonzalez Felipe,
``Phenomenological and cosmological aspects of a minimal GUT scenario,''
Nucl. Phys. B \textbf{747}, 312-327 (2006)
\arXivold{hep-ph/0512068}{hep-ph}.
%86 citations counted in INSPIRE as of 17 Nov 2023


%\cite{Dorsner:2006hw}
\bibitem{Dorsner:2006hw}
I.~Dorsner, P.~Fileviez Perez and G.~Rodrigo,
``Fermion masses and the UV cutoff of the minimal realistic SU(5),''
Phys. Rev. D \textbf{75}, 125007 (2007)
\arXivold{hep-ph/0607208}{hep-ph}.
%53 citations counted in INSPIRE as of 17 Nov 2023


\bibitem{Tran}
%\cite{Aoki:2021ckh}
%\bibitem{Aoki:2021ckh}
K.~Aoki, T.~Q.~Loc, T.~Noumi and J.~Tokuda,
%``Is the Standard Model in the Swampland? Consistency Requirements from Gravitational Scattering,''
Phys. Rev. Lett. \textbf{127} (2021) no.9, 091602
doi:10.1103/PhysRevLett.127.091602
[arXiv:2104.09682 [hep-th]] ; \\
%25 citations counted in INSPIRE as of 29 Jan 2024

%
%\cite{Loc:2023nrp}
%\bibitem{Loc:2023nrp}
T.~Q.~Loc,
%``Gravitational positivity in electroweak sector,''
[arXiv:2312.09132 [hep-th]].
%0 citations counted in INSPIRE as of 29 Jan 2024


%\cite{Raffelt:1996wa}
\bibitem{Raffelt:1996wa}
G.~G.~Raffelt,
``Stars as laboratories for fundamental physics: The astrophysics of neutrinos, axions, and other weakly interacting particles,''
1996,
ISBN 978-0-226-70272-8
%144 citations counted in INSPIRE as of 01 Oct 2023


\end{thebibliography}
\end{document}